\documentclass[12pt,a4paper]{article}
\usepackage{epsfig}
\begin{document}
\title{The branching ratio $R_{b}$ in the littlest Higgs model}

\author{Chongxing Yue and Wei Wang \\
{\small  Department of Physics, Liaoning Normal University, Dalian
116029, China}\thanks{E-mail:cxyue@lnnu.edu.cn}\\}
\date{\today}

\maketitle
\begin{abstract}
In the context of the littlest Higgs(LH) model, we study the
contributions of the new particles to the branching ratio $R_{b}$.
We find that the contributions mainly dependent on the free
parameters $f$, $c'$ and $x_{L}$. The precision measurement value
of $R_{b}$ gives severe constraints on these free parameters.

\vspace{1cm}

\end{abstract}

\newpage
\vspace{.5cm} \noindent{\bf I. Introduction}

It is well know that most of the electroweak oblique and QCD
corrections to the $Z\rightarrow b\overline{b}$ branching ratio
$R_{b}$ cancel between numerator and denominator, $R_{b}$ is very
sensitive to the new physics(NP) beyond the standard model(SM).
The precision experimental value of $R_{b}$ may give a severe
constraint on the NP[1]. Thus, it is very important to study the
$Z\rightarrow b\overline{b}$ process in extensions of the SM and
pursue the resulting implications.

Little Higgs models[2,3,4] provide a new approach to solve the
hierarchy between the $TeV$ scale of possible NP and the
electroweak scale, $\nu=246GeV=(\sqrt{2}G_{F})^{-1/2}$. In these
models, at least two interactions are needed to explicitly break
all of the global symmetries, which forbid quadratic divergences
in the Higgs mass at one-loop. Electroweak symmetry breaking(EWSB)
is triggered by a Coleman-Weinberg potential, which is generated
by integrating out the heavy degrees of freedom. In this kind of
models, the Higgs boson is a pseudo-Goldstone boson of a global
symmetry which is spontaneously broken at some higher scale $f$ by
an vacuum expectation value(VEV) and thus is naturally light. A
general feature of this kind of models is that the cancellation of
the quadratic divergences is realized between particles of the
same statistics.

Little Higgs models are weakly interaction models, which contain
extra gauge bosons, new scalars and fermions, apart from the SM
particles. These new particles might produce characteristic
signatures at the present and future collider experiments[5,6,7].
Since the extra gauge bosons can mix with the SM gauge bosons $W$
and $Z$, the masses of the SM gauge bosons $W$ and $Z$ and their
couplings to the SM particles are modified from those in the SM at
the order of $\frac{\nu^{2}}{f^{2}}$. Thus, the precision
measurement data can give severe constraints on this kind of
models[5,8,9].

Aim of this paper is to consider the $Z\rightarrow b\overline{b}$
branching ratio $R_{b}$ in the context of the littlest Higgs(LH)
model[2] and see whether the new particles predicted by the LH
model can give significant contributions to $R_{b}$. We find that,
compare the calculated value of $R_{b}$ with the experimental
measured value, the precision data can give severe constraint on
the free parameters of the LH model.

The LH model has been extensively described in literature.
However, in order to clarify notation which is relevant to our
calculation, we will simply review the LH model in section II. In
section III, we discuss the effects of the new gauge bosons on the
branching ratio $R_{b}$. We calculate the contributions of the top
quark $t$ and vector-like quark $T$ to $R_{b}$  via the couplings
$W\overline{t}b, W\overline{T}b, W'\overline{t}b$ and
$W'\overline{T}b$ in section IV. The contributions of the new
scalars to $R_{b}$ are studied in section V. Discussions and
conclusions are given in section VI.

\vspace{.5cm} \noindent{\bf II. Littlest Higgs model}

The LH model[2] is embedded into a non-linear $\sigma$ model with
the coset space of $SU(5)/SO(5)$. At the scale $\Lambda _{s}\sim
4\pi f $, the global $SU(5)$ symmetry is broken into its subgroup
$SO(5)$ via a VEV of order $f$, resulting in 14 Goldstone bosons.
The effective field theory of these Goldstone bosons is
parameterized by a non-linear $\sigma$ model with gauge symmetry
$[SU(2)\times U(1)]^{2}$, spontaneously broken down to its
diagonal subgroup $SU(2)\times U(1)$, identified as the SM
electroweak gauge group. Four of these Goldstone bosons are eaten
by the broken gauge generators, leaving 10 states that transform
under the SM gauge group as a doublet $H$ and a triplet $\Phi$. A
new charge $2/3$ quark $T$ is also needed to cancel the
divergences from the top quark loop.

The effective non-linear Lagrangian invariant under the local
gauge group $[SU(2)_{1}\times U(1)_{1}]\times [SU(2)_{2}\times
U(1)_{2}]$, which can be written as[5,9]:
\begin{equation}
\pounds_{eff}=\pounds_{G}+\pounds_{F}+\pounds_{Y}+\pounds_{\Sigma}-V_{CW},
\end{equation}
where $\pounds_{G}$ consists of the pure gauge terms, which can
give the $3-$ and $4-$particle interactions among the $SU(2)$
gauge bosons and the couplings of the $U(1)$ gauge bosons to the
$SU(2)$ gauge bosons. The fermion kinetic term $\pounds_{F}$ can
give the couplings of the gauge bosons to fermions. The couplings
of the scalars $H$ and $\Phi$ to fermions can be derived from the
Yukawa interaction term $\pounds_{Y}$. In the LH model, the global
symmetry prevents the appearance of a Higgs potential at tree
level. The effective Higgs potential, the Coleman-Weinberg
potential $V_{CW}$[10], is generated at one-loop and higher orders
due to interactions with gauge bosons and fermions, which can
induce to EWSB by driving the Higgs mass squared parameter
negative. $\pounds_{\Sigma}$ consists of the $\sigma$ model terms
of the LH model. The scalar $\Sigma$ field can be written as:
\begin{equation}
\Sigma =e^{i\Pi/f}\langle \Sigma_{0} \rangle
e^{i\Pi^{T}/f}=e^{2i\Pi/f}\langle \Sigma_{0} \rangle
\end{equation}
with $\langle \Sigma_{0} \rangle \sim f$ which generates masses
and mixing between the gauge bosons. The ten pseudo-Goldstone
bosons can be parameterized as:
\begin{equation}
\Pi = \left( \begin{array}{ccc}0&H^{+}/\sqrt{2}&\Phi^{+}\\
H/\sqrt{2}&0&H^{\ast}/\sqrt{2}\\\Phi&H^{T}/\sqrt{2}&0\end{array}\right).
\end{equation}
Where $H$ is identified as the SM Higgs doublet,
$H=(H^{+},H^{0})$, and $\Phi$ is a complex $SU(2)$ triplet with
hypercharge $Y=2$,
\begin{equation}
\Phi = \left( \begin{array}{cc}\Phi^{++}&\Phi^{+}/\sqrt{2}\\
\Phi^{+}/\sqrt{2}&\Phi^{0}\end{array}\right).
\end{equation}
The kinetic terms of the scalar $\Sigma$ field are given by
\begin{equation}
\pounds_{\Sigma}=\frac{f^{2}}{8}Tr\{(D_{\mu}\Sigma)(D^{\mu}\Sigma)^{+}\},
\end{equation}
where the covariant derivative of the $\Sigma$ field is defined
as:
\begin{equation}
D_{\mu}\Sigma=\partial_{\mu}\Sigma-i\sum_{j}[g_{j}W^{a}_{j}(Q^{a}_{j}\Sigma+\Sigma
Q^{aT}_{j})+g_{j}'B_{j}(Y_{j}\Sigma+\Sigma Y_{j}^{T})],
\end{equation}
where $g_{j}, g'_{j}$ are the gauge coupling constants, $W_{j},
B_{j}$ are the gauge bosons, $Q_{j}^{a}$ and $Y_{j}$ are the
generators of gauge transformations. The gauge boson mass
eigen-states are given by
\begin{equation}
\begin{array}{cc}W=sW_{1}+cW_{2},&W'=-cW_{1}+sW_{2},\\
B=s'B_{1}+c'B_{2},&B'=-c'B_{1}+s'B_{2}\end{array}
\end{equation}
with the cosines of two mixing angles,
\begin{equation}
c=\frac{g_{1}}{\sqrt{g_{1}^{2}+g_{2}^{2}}},\hspace{2cm}c'=\frac{g'_{1}}{\sqrt{g'^{2}_{1}+g'^{2}_{2}}}.
\end{equation}
The SM gauge coupling constants are $g=g_{1}s=g_{2}c$ and
$g'=g'_{1}s'=g'_{2}c'$.

From the effective non-linear Lagrangian ${\cal L}$, one can
derive the mass and coupling expressions of the gauge bosons,
scalars and the fermions, which have been extensively discussed in
Refs.[5,9]. The mass spectrum of the LH model and the coupling
forms which are related our calculations are summarized in
appendix A , B and C, respectively.

\vspace{.5cm} \noindent{\bf III. New gauge bosons and the
$Z\rightarrow b \overline{b}$ branching ratio $R_{b}$}

1. Corrections of new physics to the $Z\rightarrow b\overline{b}$
branching ratio $R_{b}$

In general, the effective $Z\rightarrow b\overline{b}$ vertex can
be written as:
\begin{equation}
[g_{L}^{b}\overline{b}_{L}\gamma^{\mu}
b_{L}+g_{R}^{b}\overline{b}_{R}\gamma^{\mu} b_{R}]\cdot Z^{\mu}
\end{equation}
with the form factors $g_{L}^{b}$ and $g_{R}^{b}$:

\begin{eqnarray}
g_{L}^{b}&=& g_{L}^{b,SM}+\delta g_{L}^{b}= \frac{e}{S_{W}C_{W}}(-\frac{1}{2}+\frac{1}{3}
S_{W}^{2})+\delta g_{L}^{b},\nonumber\\
g_{R}^{b}&=& g_{R}^{b,SM}+\delta g_{R}^{b}=
\frac{e}{S_{W}C_{W}}(\frac{1}{3}S_{W}^{2})+\delta g_{R}^{b}.
\end{eqnarray}
Where $S_{W}=\sin \theta_{W}$, $\theta_{W}$ is the Weinberg angle.
$\delta g_{L}^{b}$ and $\delta g_{R}^{b}$ represent the
corrections of NP to the left-handed and right-handed
$Zb\overline{b}$ couplings, respectively. Certainly, the
corrections of NP to the $Z b\overline{b}$ couplings $g_{L}^{b}$
and $g_{R}^{b}$ may give rise to one additional form factor,
proportional to $\sigma^{\mu\nu}q^{\nu}$. However, its
contributions to $R_{b}$ are very small and can be ignored[1].

The branching ratio $R_{b}$ can be written as:
\begin{equation}
R_{b}=\frac{\Gamma_{b}}{\Gamma_{h}}=\frac{\Gamma_{b}}{3\Gamma_{b}+2\Gamma_{c}}.
\end{equation}
Here $\Gamma_{c}$ is the width of the process $Z\rightarrow c
\overline{c}$. The partial decay width, $\Gamma_{q}$, of the
$Z\rightarrow q \overline{q}$ decay($q=u, d, c, s$, and $b$) is
given as[11]:
\begin{equation}
\Gamma_{q}=6\Gamma_{0}(1+\frac{\alpha_{s}}{\pi})[(g_{L}^{q})^{2}+(g_{R}^{q})^{2}],
\end{equation}
with $\Gamma_{0}=\frac{G_{F}m_{Z}^{3}}{24\sqrt{2}\pi}$. The factor
$\frac{\alpha_{s}}{\pi}$ contain contributions from the find state
gluons and photons. In above equation, the masses of the final
quarks are assumed to be negligible.

Since the branching ratio $R_{b}$ is the ratio between two
hadronic widths, $R_{b}$ is almost independent of the EW oblique
and QCD corrections because of the near cancellation of these
corrections between the numerator and the denominator. The
remaining ones are absorbed in the definition of the renormalized
coupling parameters $\alpha$ and $S_{W}$, up to terms of high
order in the electroweak corrections[12]. Thus, $R_{b}$ is very
sensitive to the NP beyond the SM. The correction of NP to $R_{b}$
can be written as:
\begin{eqnarray}
\delta R_{b}&=&R_{b}-R_{b}^{SM}=\frac{\Gamma_{b}^{SM}+\delta
\Gamma_{b}}{\Gamma_{h}^{SM}+\delta
\Gamma_{h}}-\frac{\Gamma_{b}^{SM}}{\Gamma_{h}^{SM}}\approx
(\frac{\delta \Gamma_{b}}{\Gamma_{b}}-\frac{\delta
\Gamma_{h}}{\Gamma_{h}})R_{b}^{SM}\nonumber\\
&=&R_{b}^{SM}\{\frac{2(g_{L}^{b}\delta g_{L}^{b}+g_{R}^{b}\delta
g_{R}^{b})}{(g_{L}^{b})^{2}+(g_{R}^{b})^{2}}-\frac{4(g_{L}^{c}\delta
g_{L}^{c}+g_{R}^{c}\delta g_{R}^{c})+6(g_{L}^{b}\delta
g_{L}^{b}+g_{R}^{b}\delta
g_{R}^{b})}{2[(g_{L}^{c})^{2}+(g_{R}^{c})^{2}]+3[(g_{L}^{b})^{2}+(g_{R}^{b})^{2}]}\}.
\end{eqnarray}
In above equation, we have neglected the terms of $O[(\delta
g_{L,R}^{q} )^{2}]$. In the next subsection, we will study the
corrections of the new gauge bosons predicted by the LH model to
the branching ratio $R_{b}$.

2. The extra gauge bosons and the branching ratio $R_{b}$

The LH model predicts the existence of the extra gauge bosons,
such as $W', Z'$ and $B'$. These new particles can generate
corrections to the branching ratio $R_{b}$ via mixing with the SM
gauge bosons and the coupling to the SM fermions. The corrections
to $R_{b}$ mainly come from three sources: (1) the modifications
of the relations between the SM parameters and the precision
electroweak input parameters, which come from the mixing of the
heavy $W'$ boson to the couplings of the charge current and from
the contributions of the current to the equations of motion of the
heavy gauge bosons, (2) the correction terms to the
$Zb\overline{b}$ couplings $g_{L}^{b}$ and $g_{R}^{b}$ coming from
the mixing between the extra gauge boson $Z'$ and the SM gauge
boson $Z$, (3) the neutral gauge bosons $Z'$ exchange and $B'$
exchange. In the LH model, the relation between the Fermi coupling
constant $G_{F}$, the gauge boson $Z$ mass $m_{Z}$ and the fine
structure coupling constant $\alpha(m_{Z})$ can be written as[9]:
\begin{equation}
\frac{G_{F}}{\sqrt{2}}=\frac{\pi\alpha}{2\sqrt{2}m_{Z}^{2}S_{W}^{2}C_{W}^{2}}[1-\frac{g}{G_{F}}
\frac{c}{s}(c^{2}-s^{2})\frac{\nu^{2}}{f^{2}}+2c^{4}\frac{\nu^{2}}{f^{2}}-\frac{5}{4}(c'^{2}-
s'^{2})^{2}\frac{\nu^{2}}{f^{2}}].
\end{equation}
So we have
\begin{equation}
\frac{e^{2}}{S_{W}^{2}C_{W}^{2}}=\frac{8G_{F}m^{2}_{Z}}{1-\frac{g}{G_{F}}\frac{c}{s}(c^{2}-s^{2})
\frac{\nu^{2}}{f^{2}}+2c^{4}\frac{\nu^{2}}{f^{2}}-\frac{5}{4}(c'^{2}-s'^{2})^{2}\frac{\nu^{2}}{f^{2}}}.
\end{equation}
In our numerical calculations, we will take $G_{F}=1.16637\times
10^{-5}GeV^{-2}$, $m_{Z}=91.187GeV$ and $m_{t}=174.3GeV$ [13] as
input parameters and use them to represent the other SM
parameters.

Due to the mixing between the gauge bosons $Z$ and $Z'$, the
tree-level $Zq\overline{q}$ couplings $g_{L}^{q,SM}$ and
$g_{R}^{q,SM}$ receive corrections at the order of
$\frac{\nu^{2}}{f^{2}}$:
\begin{eqnarray}
\delta
g_{L}^{q_{i},1}&=&\frac{e}{S_{W}C_{W}}\frac{\nu^{2}}{f^{2}}[\frac{c^{2}(c^{2}-s^{2})}{4}+\frac{5}{6}
(c'^{2}-s'^{2})(-\frac{1}{5}+\frac{1}{2}c'^{2})],\\
\delta
g_{R}^{q_{i},1}&=&\frac{e}{S_{W}C_{W}}\frac{\nu^{2}}{f^{2}}[\frac{5}{3}(c'^{2}-s'^{2})
(\frac{1}{5}-\frac{1}{2}c'^{2})],\\
\delta
g_{L}^{q_{j},1}&=&\frac{e}{S_{W}C_{W}}\frac{\nu^{2}}{f^{2}}[-\frac{c^{2}(c^{2}-s^{2})}{4}+\frac{5}{6}
(c'^{2}-s'^{2})(\frac{4}{5}-\frac{1}{2}c'^{2})],\\
\delta
g_{R}^{q_{j},1}&=&\frac{e}{S_{W}C_{W}}\frac{\nu^{2}}{f^{2}}[\frac{5}{3}(c'^{2}-s'^{2})
(\frac{1}{10}+\frac{1}{2}c'^{2})],
\end{eqnarray}
where $q_{i}$ and $q_{j}$ represent the down-type quarks($d, s,
b$) and the up-type quarks(c, s),respectively.

\begin{figure}[htb]
\vspace{-0.5cm}
\begin{center}
\epsfig{file=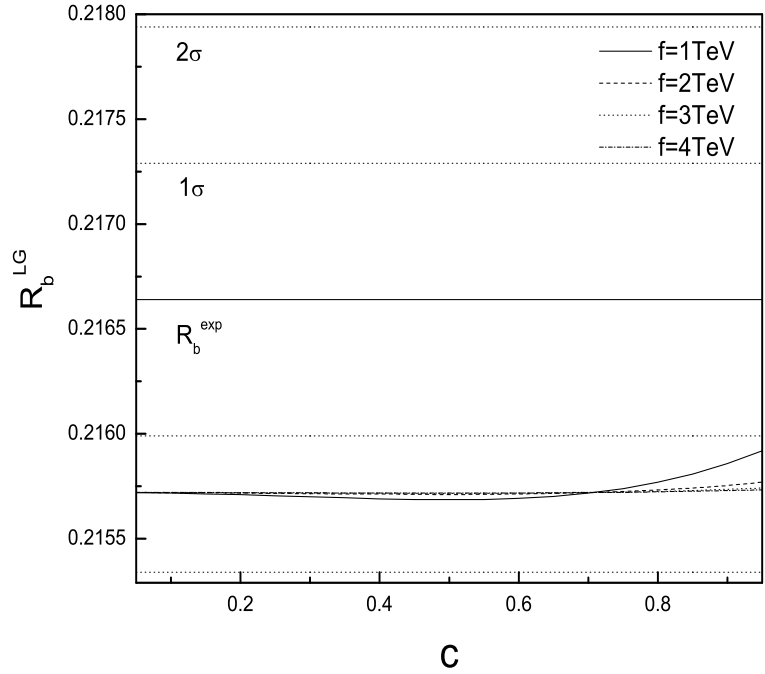,width=320pt,height=300pt} \vspace{-1.0cm}
\hspace{5mm} \caption{The branching ratio $R_{b}^{LG}$ as a
function of the mixing parameter $c$ for the mixing parameter
$c'=\frac{1}{\sqrt{2}}$, $f=1TeV$, $2TeV$, $3TeV$ and $4TeV$.}
 \label{ee}
\end{center}
\end{figure}

Using the same method as calculating the contributions of the
topcolor gauge bosons to $R_{b}$[14,15], we can give the
corrections of the neutral gauge boson $Z'$ exchange and $B'$
exchange to the $Zq\overline{q}$ couplings $g_{L}^{q}$ and
$g_{R}^{q}$:
\begin{eqnarray}
\delta
g_{L}^{q_{i},2}&=&\frac{e^{2}c^{2}}{24\pi^{2}S^{2}_{W}s^{2}}\frac{m_{Z}^{2}}
{M_{Z'}^{2}}\ln\frac{M_{Z'}^{2}}
{m_{Z}^{2}}g_{L}^{q_{i},SM},\hspace{0.5cm}\delta
g_{R}^{q_{i},2}=0,\\
\delta
g_{L}^{q_{j},2}&=&\frac{e^{2}c^{2}}{24\pi^{2}S^{2}_{W}s^{2}}\frac{m_{Z}^{2}}
{M_{Z'}^{2}}\ln\frac{M_{Z'}^{2}}
{m_{Z}^{2}}g_{L}^{q_{j},SM},\hspace{0.5cm}\delta
g_{R}^{q_{j},2}=0,\\
\delta
g_{L}^{q_{i},3}&=&\frac{e^{2}}{54\pi^{2}C^{2}_{W}s'^{2}c'^{2}}[\frac{1}{5}
-\frac{1}{2}c'^{2}]^{2}\frac{m_{Z}^{2}}
{M_{B'}^{2}}\ln\frac{M_{B'}^{2}}{m_{Z}^{2}}g_{R}^{q_{i},SM},\\
\delta
g_{R}^{q_{i},3}&=&\frac{2e^{2}}{27\pi^{2}C^{2}_{W}s'^{2}c'^{2}}[-\frac{1}{5}+\frac{1}{2}c'^{2}]^{2}
\frac{m_{Z}^{2}}{M_{B'}^{2}}\ln\frac{M_{B'}^{2}}{m_{Z}^{2}}g_{R}^{q_{i},SM},\\
\delta
g_{L}^{q_{j},3}&=&\frac{e^{2}}{54\pi^{2}C^{2}_{W}s'^{2}c'^{2}}[\frac{1}{5}
-\frac{1}{2}c'^{2}]^{2}\frac{m_{Z}^{2}}
{M_{B'}^{2}}\ln\frac{M_{B'}^{2}}{m_{Z}^{2}}g_{R}^{q_{j},SM},\\
\delta
g_{R}^{q_{j},3}&=&\frac{8e^{2}}{27\pi^{2}C^{2}_{W}s'^{2}c'^{2}}[\frac{1}{5}-\frac{1}{2}c'^{2}]^{2}
\frac{m_{Z}^{2}}{M_{B'}^{2}}\ln\frac{M_{B'}^{2}}{m_{Z}^{2}}g_{R}^{q_{j},SM}.
\end{eqnarray}
Where $M_{Z'}$ and $M_{B'}$ are the masses of the gauge boson $Z'$
and the heavy photon $B'$, respectively, which have been listed in
appendix A. In above equations, we have used the expressions of
the $Z'q\overline{q}$ and $B'q\overline{q}$ couplings given in
appendix B. Adding all the corrections together, we obtain the
total corrections of extra gauge bosons to $R_{b}$:
\begin{eqnarray}
\delta g_{L}^{b,G}&=&\delta g_{L}^{b,1}+\delta g_{L}^{b,2}+\delta
g_{L}^{b,3}, \hspace{1cm}\delta g_{R}^{b,G}=\delta
g_{R}^{b,1}+\delta g_{R}^{b,3},\\
\delta g_{L}^{c,G}&=&\delta g_{L}^{c,1}+\delta g_{L}^{c,2}+\delta
g_{L}^{c,3}, \hspace{1cm}\delta g_{R}^{c,G}=\delta
g_{R}^{c,1}+\delta g_{R}^{c,2}+\delta g_{R}^{c,3}.
\end{eqnarray}

Plugging Eqs.(15)-(27) into Eq.(13), we can obtain the relative
correction $\frac{\delta R_{b}^{LG}}{R_{b}^{SM}}$ given by the new
gauge bosons. In our calculation, we have taken
$R_{b}^{LG}=R_{b}^{SM}+\delta R_{b}^{LG}$, $R_{b}^{SM}=0.21572$,
and $R_{b}^{exp}=0.21664\pm 0.00065$[16]. Our numerical results
are summarized in Fig.1 and Fig.2, in which we have used the
horizontal solid line to denote the central value of $R_{b}^{exp}$
and dotted lines to show the $1\sigma$ and $2\sigma$ bounds. In
Fig.1(Fig.2) we plot $R_{b}^{LG}$ as a function of the mixing
parameter $c(c')$ for
$c'=\frac{1}{\sqrt{2}}(c=\frac{1}{\sqrt{2}})$ and the scale
parameter $f=1TeV$(solid line), $2TeV$(dashed line), $3TeV$(dotted
line) and $4TeV$(dotted-dashed line). From Fig.1 and Fig.2, one
can see that the contributions of the new gauge bosons to $R_{b}$
decrease as the scale parameter $f$ increasing. For
$c'=\frac{1}{\sqrt{2}}$, $R_{b}^{LG}$ is insensitive to the mixing
parameter $c$ and the value of $\delta R_{b}^{LG}$ is very small.
For $c=\frac{1}{\sqrt{2}}$, the new gauge bosons decrease the
value of the branching ratio $R_{b}$ for $c'> 0.72$ and $f>1TeV$.
To make the predicted $R_{b}$ value satisfy the precision
experimental value in $2\sigma$ bound, we should have
$0.57<c'<0.73$ for $f=1TeV$. For $c=\frac{1}{\sqrt{2}}$, $f>2TeV$,
the predicted value of $R_{b}$ is consistent with the precision
experimental value $R_{b}^{exp}$ within $2\sigma$ bound in most of
the allowed range of the mixing parameter $c'$.

\begin{figure}[htb]
\vspace{-0.5cm}
\begin{center}
\epsfig{file=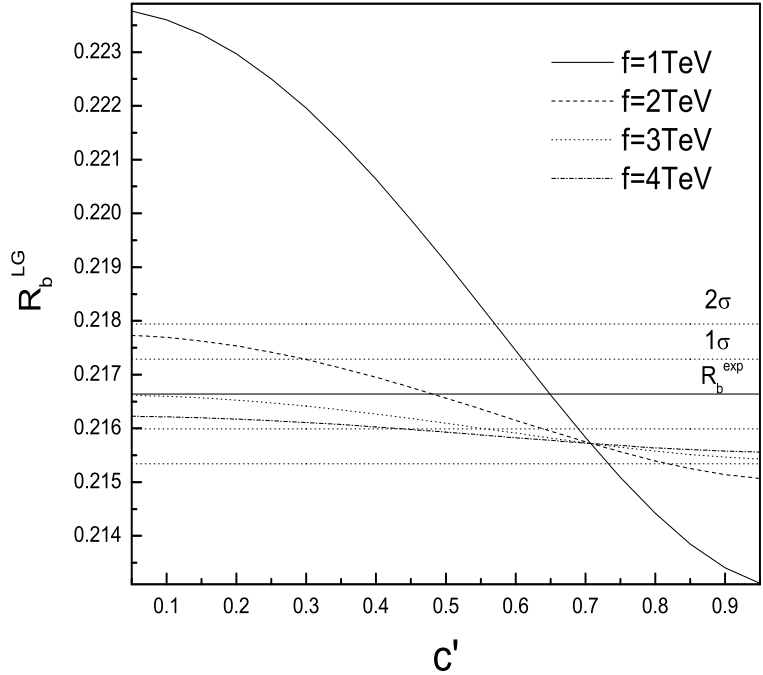,width=320pt,height=300pt} \vspace{-1cm}
\hspace{5mm} \caption{The branching ratio $R_{b}^{LG}$ as a
function of the mixing parameter $c'$ for $c=\frac{1}{\sqrt{2}}$
and $f=1TeV$, $2TeV$, $3TeV$ and $4TeV$.}
 \label{ee}
\end{center}
\end{figure}

From above discussions, we can see that the corrections of new
gauge bosons to $R_{b}$ can be divided into two parts: one part is
the tree-level corrections coming from the shift in the $Z$
couplings to quarks and the modifications of the relations between
the SM parameters and the precision electroweak input parameters
and the second part is the one-loop corrections coming from the
neutral gauge bosons $Z'$ exchange and $B'$ exchange. To compare
the tree-level corrections with the one-loop corrections, we plot
$R=|\delta R_{b}^{1-loop}/\delta R_{b}^{tree-level}|$ as a
function of the mixing parameter $c'$ for $f=2TeV$ and
$c=\frac{1}{\sqrt{2}}$ in Fig.3. One can see from Fig.3 that the
one-loop contribution is smaller than the tree-level contribution
at least by two orders of magnitude in all of the parameter space.

\begin{figure}[htb]
\vspace{0cm}
\begin{center}
\epsfig{file=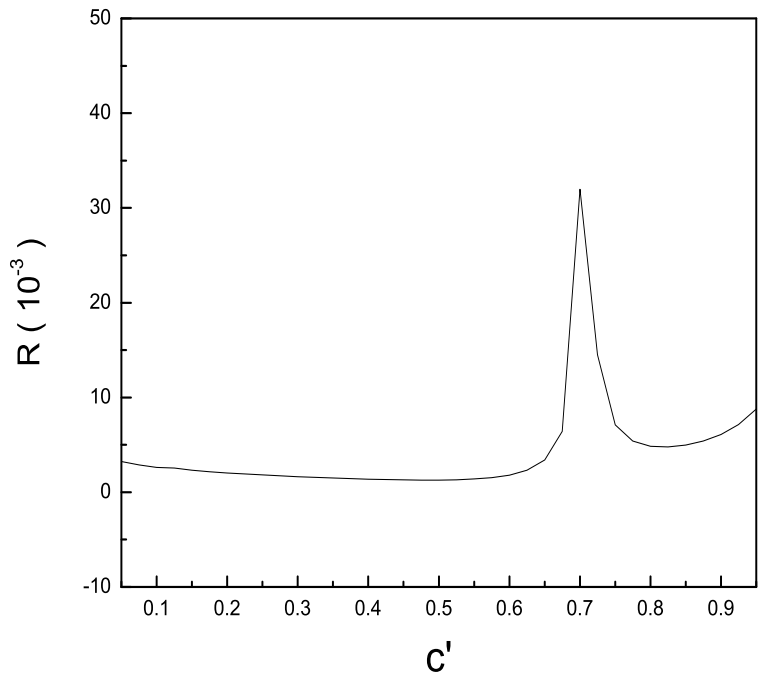,width=320pt,height=300pt} \vspace{-1.5cm}
\hspace{5mm} \caption{The relative correction $R$ as a function of
the mixing parameter $c'$ for $f=2TeV$ and
 $c=\frac{1}{\sqrt{2}}$.} \label{ee}
\end{center}
\end{figure}

\vspace{0.2cm} \noindent{\bf IV. The corrections of the top quark
$t$ and vector-like quark $T$ to $R_{b}$}

It is well known that $R_{b}$ is almost independent of the EW
oblique and QCD corrections. However, for $Zb\overline{b}$
couplings, there is an important correction coming from the top
triangle diagrams, which can not be ignored. They can generate
significant $m_{t}-$enhanced contributions to $R_{b}$[17].
Furthermore, the extra top quark predicted by NP can also produce
significant corrections to $R_{b}$ at one-loop[1,18]. In this
section, we will calculate the corrections of the top quark $t$
and vector-like quark $T$ to $R_{b}$ via the couplings
$W\overline{t}b$, $W\overline{T}b$, $W'\overline{t}b$ and
$W\overline{T}b$. The relevant Feynman diagrams are shown in
Fig.4.

Since the gauge bosons $W$ and $W'$ can only couple to the left
handed quark $s, t, T$ and $b$, the top and vector-like top
triangle loops have no contributions to the right-handed
$Zb\overline{b}$ coupling $g_{R}^{b}$. If we assume that the mass
of the bottom quark is approximately equal to zero, then the
corrections to the $Zb\overline{b}$ coupling $g_{L}^{b}$ generated
by the $W\overline{t}b$ and $W'\overline{t}b$ couplings can be
written as:
\begin{equation}
\delta g_{Lt}^{b,1}=(\frac{e}{S_{W}C_{W}})\{-\frac{\alpha}{4\pi
S_{W}^{2}}[F_{1}(x_{t})+\frac{c^{2}}{s^{2}}F_{1}(x'_{t})]+\frac{3\alpha
C_{W}^{2}}{8\pi
S^{2}_{W}}[F_{2}(x_{t})+\frac{c^{2}}{s^{2}}F_{2}(x'_{t})]\}
\end{equation}
with
\begin{eqnarray}
F_{1}(x)&=&\frac{g_{L}^{t}}{2}[\frac{x(x-2)}{(x-1)^{2}}\ln x
+\frac{x}{x-1}]+g_{R}^{t}[\frac{x}{(x-1)^{2}}\ln x-\frac{x}{x-1}],\\
F_{2}(x)&=&\frac{x^{2}}{(x-1)^{2}}\ln x-\frac{x}{x-1},
\end{eqnarray}
where $x_{t}=(\frac{m_{t}}{m_{W}})^{2}$ and
$x'_{t}=(\frac{m_{t}}{M_{W'}})^{2}$. In above equation, we have
neglected the interference effects between gauge bosons $W'$ and
$W$.

\begin{figure}[htb]
\vspace{-7.5cm}
\begin{center}
\epsfig{file=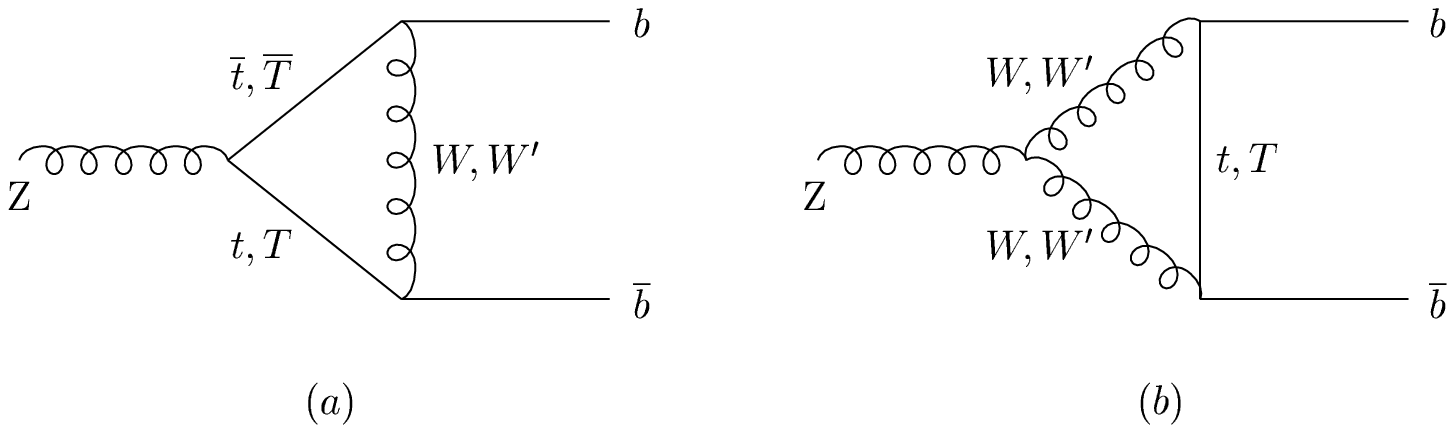,width=500pt,height=650pt} \vspace{-10cm}
\hspace{5mm} \caption{Feynman diagrams for the contributions of
the top quark $t$ and vector-like quark $T$ to the
$Zb\overline{b}$ vertex .} \label{ee}
\end{center}
\end{figure}

In the LH model, due to the mixing between the top quark $t$ and
the vector like quark $T$, the tree-level $Zt\overline{t}$
couplings receive corrections at the order of
$\frac{\nu^{2}}{f^{2}}$, which also have contributions to the
$Zb\overline{b}$ coupling $g_{L}^{b}$:
\begin{equation}
\delta g_{Lt}^{b,2}=-(\frac{e}{S_{W}C_{W}})\frac{\alpha}{16\pi
S_{W}^{2}}(\frac{\nu^{2}x_{L}^{2}}{f^{2}})[x_{t}(2-\frac{4}{x_{t}-1}\log
x_{t})+\frac{c^{2}}{s^{2}}x'_{t}(2-\frac{4}{x'_{t}-1}\log
x'_{t})].
\end{equation}
The mixing angle parameter between the SM top quark $t$ and the
vector-like quark $T$ is defined as
$x_{L}=\frac{\lambda_{1}^{2}}{\lambda_{1}^{2}+\lambda_{2}^{2}}$,
in which  $\lambda_{1}$ and $\lambda_{2}$ are the coupling
parameters.

\begin{figure}[htb]
\vspace{-0.5cm}
\begin{center}
\epsfig{file=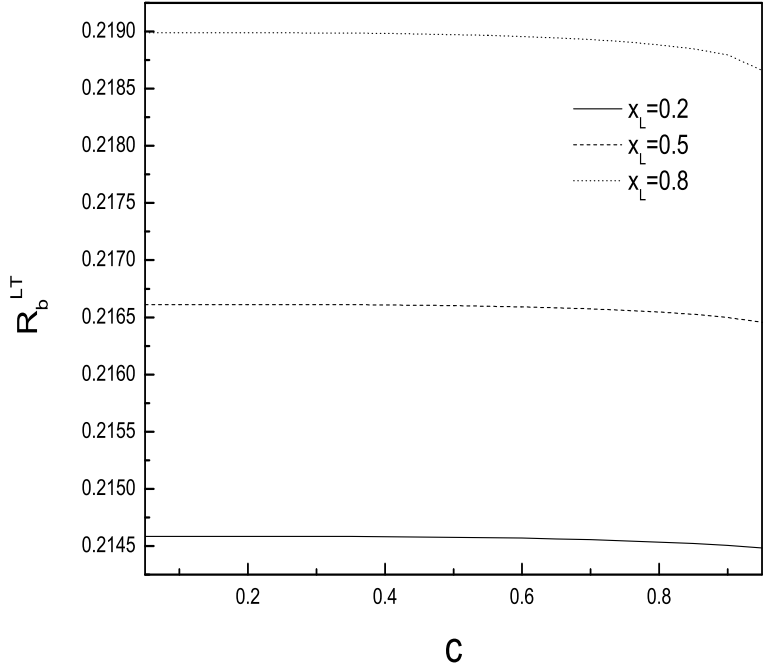,width=320pt,height=300pt} \vspace{-1cm}
\hspace{5mm} \caption{The branching ratio $R_{b}^{LT}$ as a
function of the mixing parameter $c$ for $f=2TeV$ and three values
of the scale parameter $x_{L}$.}
 \label{ee}
\end{center}
\end{figure}

The contributions of the vector like top quark $T$ to $R_{b}$ via
the couplings $W\overline{T}b$ and $W'\overline{T}b$ can be
written as:
\begin{equation}
\delta
g_{Lt}^{b,3}=(\frac{e}{S_{W}C_{W}})\frac{\nu^{2}x_{L}^{2}}{f^{2}}\{-\frac{\alpha}{4\pi
S_{W}^{2}}[F_{3}(x_{T})+\frac{c^{2}}{s^{2}}F_{3}(x'_{T})]+\frac{3\alpha
C_{W}^{2}}{8\pi
S^{2}_{W}}[F_{2}(x_{T})+\frac{c^{2}}{s^{2}}F_{2}(x'_{T})]\}
\end{equation}
with
\begin{equation}
F_{3}(x)=\frac{g_{L}^{T}}{2}[\frac{x(x-2)}{(x-1)^{2}}\ln x
+\frac{x}{x-1}]+g_{R}^{T}[\frac{x}{(x-1)^{2}}\ln x-\frac{x}{x-1}],
\end{equation}
where $x_{T}=(\frac{M_{T}}{m_{W}})^{2}$ and
$x'_{T}=(\frac{M_{T}}{M_{W'}})^{2}$. The $t-T$ contributions can
be given by
\begin{eqnarray}
\delta g_{Lt}^{b,4}&=&(\frac{e}{S_{W}C_{W}})\frac{\alpha}{4\pi
S_{W}^{2}}(\frac{\nu
x_{L}}{f})[\frac{1}{x_{T}-x_{t}}(\frac{x_{T}^{2}}{x_{T}-1}\log
x_{T}-\frac{x_{t}^{2}}{x_{t}-1}\log
x_{t})\nonumber\\
&&-\frac{x_{t}x_{T}}{x_{T}-x_{t}}(\frac{x_{T}}{x_{T}-1}\log
x_{T}-\frac{x_{t}}{x_{t}-1}\log x_{t})].
\end{eqnarray}

Being CKM suppression, the contributions of the top quark $t$ and
vector-like quark $T$ to the couplings of the gauge boson $Z$ to
other quarks($u, c, d, s$) are very small, which can be ignored.
Then Eq.(13) should be changed as, for calculating the
contributions of the quarks $t$ and $T$:
\begin{equation}
\delta R_{b}=2R_{b}^{SM}\{\frac{g_{L}^{b}\delta
g_{L}^{b}+g_{R}^{b}\delta
g_{R}^{b}}{(g_{L}^{b})^{2}+(g_{R}^{b})^{2}}-\frac{g_{L}^{b}\delta
g_{L}^{b}+g_{R}^{b}\delta
g_{R}^{b}}{2[(g_{L}^{c})^{2}+(g_{R}^{c})^{2}]+3[(g_{L}^{b})^{2}+(g_{R}^{b})^{2}]}\}.
\end{equation}

\begin{figure}[htb]
\vspace{-0.5cm}
\begin{center}
\epsfig{file=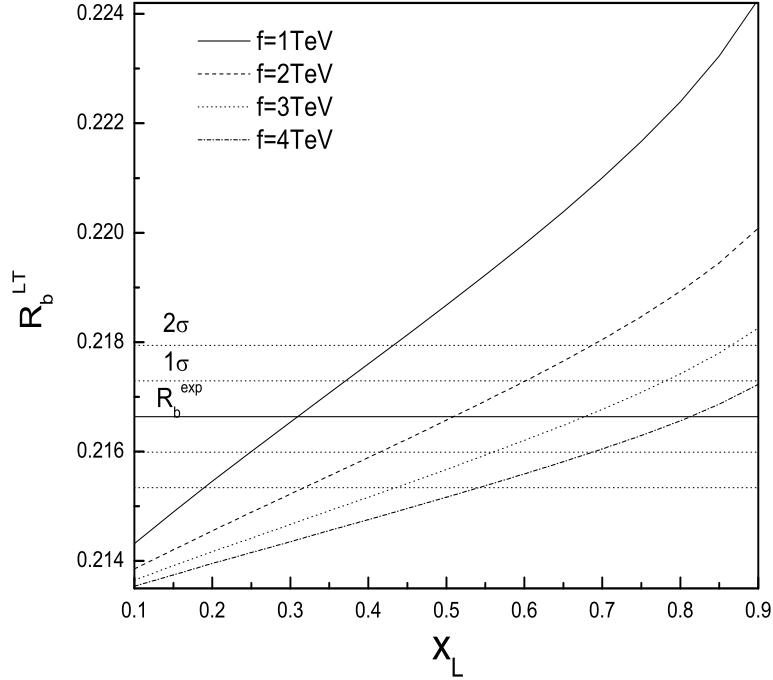,width=320pt,height=300pt} \vspace{-1cm}
\hspace{5mm} \caption{The branching ratio $R_{b}^{LT}$ as a
function of the mixing parameter $x_{L}$ for
$c=\frac{1}{\sqrt{2}}$ and four values of the scale parameter
$f$.}
 \label{ee}
\end{center}
\end{figure}

In Fig.5 we plot the branching ratio $R_{b}^{LT}=R_{b}^{SM}+\delta
R_{b}^{t}+\delta R_{b}^{T}$ as a function of the mixing parameter
$c$ for $f=2TeV$ and three values of the mixing parameter $x_{L}$.
One can see from Fig.5 that the corrections of the top quark $t$
and vector-like quark $T$ to $R_{b}$ are not sensitive to the
value of the flavor mixing parameter $c$, while are strongly
dependent on the mixing parameter $x_{L}$. This is because the
mass $M_{W'}$ of the heavy gauge boson $W'$ suppresses the
contributions of the $t$ and $T$ quarks to $R_{b}$. To see the
effects of $x_{L}$ varying on $R_{b}$, we plot $R_{b}^{LT}$ as a
function of $x_{L}$ for $c=\frac{1}{\sqrt{2}}$, and four values of
the scale parameter $f$ in Fig.6. One can see from Fig.6 that the
top quark $t$ and vector-like quark $T$ can generate negative
corrections to $R_{b}$ for $f\geq 1TeV$ and $x_{L}\leq 0.25$,
while they can give positive corrections to $R_{b}$ for $f\leq
4TeV$ and $x_{L}\geq 0.66$.

\vspace{0.5cm} \noindent{\bf V. The contributions of scalars to
the branching ratio $R_{b}$}

\begin{figure}[htb]
\vspace{-1.5cm}
\begin{center}
\epsfig{file=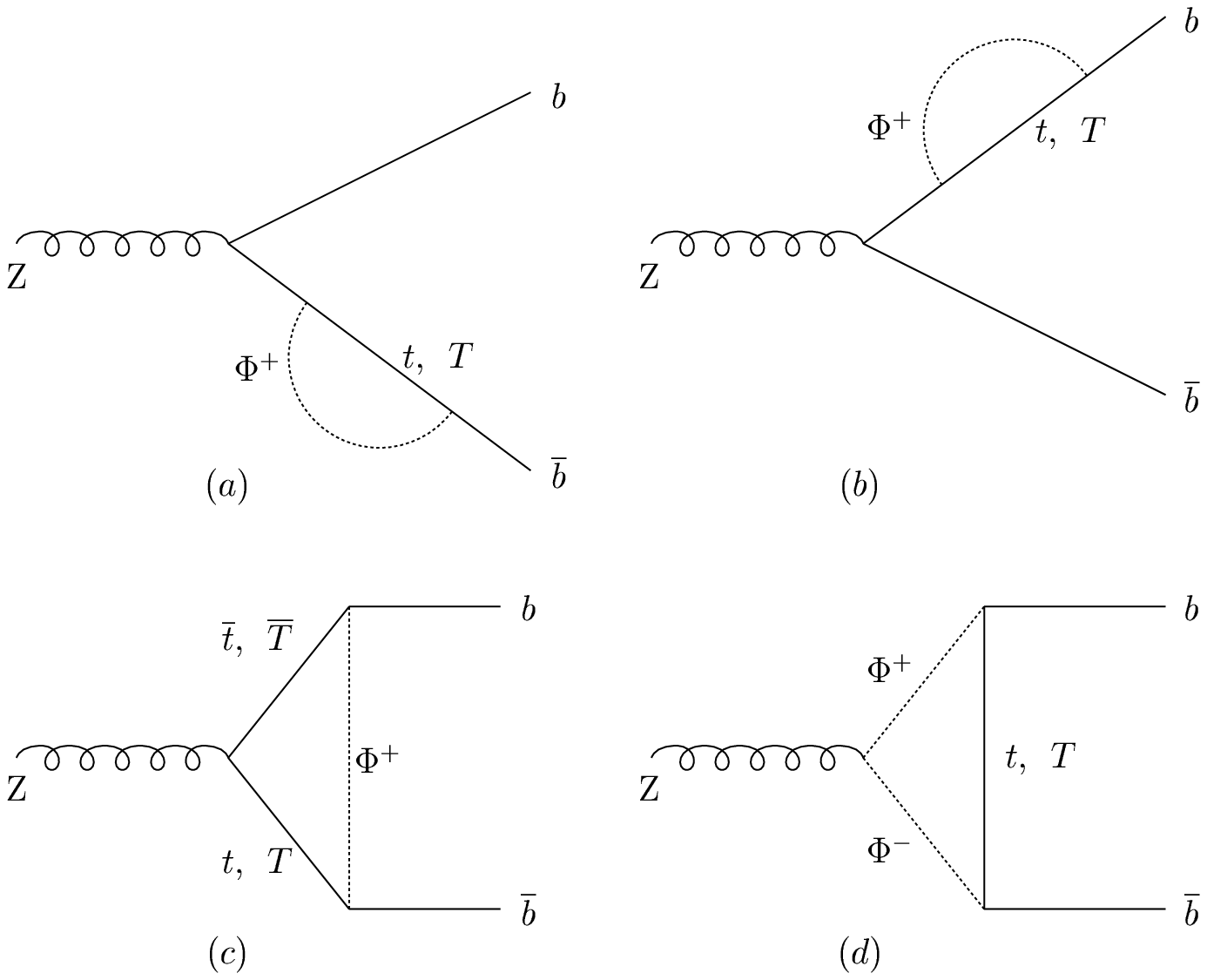,width=450pt,height=570pt} \vspace{-11.5cm}
\hspace{5mm} \caption{Feynman diagrams for the contributions of
the charged scalars $\Phi^{\pm}$ to the $Zb\overline{b}$ vertex
via the couplings $\Phi\overline{t}b$ and $\Phi\overline{T}b$.}
\label{ee}
\end{center}
\end{figure}

Since the doubly charged scalars can not couple to the SM
fermions, so they have no contributions to $R_{b}$. The singly
charged scalars $\Phi^{\pm}$ can give contributions to the
branching ratio $R_{b}$ via the couplings $\Phi\overline{t}b$ and
$\Phi\overline{T}b$. The relevant Feynman diagrams for the
corrections of the charged scalars $\Phi^{\pm}$ to the $Z
b\overline{b}$ couplings $g_{L}^{b}$ and $g_{R}^{b}$ are shown in
Fig.7.

Using the Feynman rules given in appendix C and other Feynman
rules, we can give:
\begin{eqnarray}
\delta g_{R}^{b,s}&=&0, \nonumber\\
\delta
g_{L}^{b,s}&=&\frac{e}{S_{W}C_{W}}\frac{m_{t}^{2}}{32\pi^{2}\nu^{2}}
(\frac{\nu}{f}-4\frac{\nu'}{\nu})[A_{t}+\frac{x_{L}}
{1-x_{L}}A_{T}],
\end{eqnarray}
with
\begin{eqnarray}
A_{q}&=&-g_{L}^{b,SM}\overline{B_{1}}(-P_{b},m_{q},M_{\Phi})+g_{R}^{t,SM}[2\overline{C}_{24}^{\ast}
(P_{b},-k,M_{\Phi},m_{q},m_{q})+\overline{B_{0}}(-k,m_{f},m_{q})\nonumber\\
&&\hspace{1cm}-M^{2}_{\Phi}C_{0}^{\ast}(P_{b},-k,M_{\Phi},m_{q},m_{q})]+m^{2}_{t}
g_{L}^{t,SM}C_{0}^{\ast}(P_{b},-k,M_{\Phi},m_{q},m_{q})\nonumber\\
&&\hspace{1cm}+s_{W}^{2}\overline{C}_{24}(-P_{b},k,m_{q},M_{\Phi},M_{\Phi}),
\end{eqnarray}
where $q$ represents the SM top quark $t$ or the vector-like quark
$T$. $B_{i}$, $C_{i}$ and $C_{ij}$ are the standard Feynman
integrals[19], in which the variable $P_{b}$ is the momentum of
$b$ quark, $k$ is the momentum of the gauge boson $Z$.

\begin{figure}[htb]
\vspace{-0.5cm}
\begin{center}
\epsfig{file=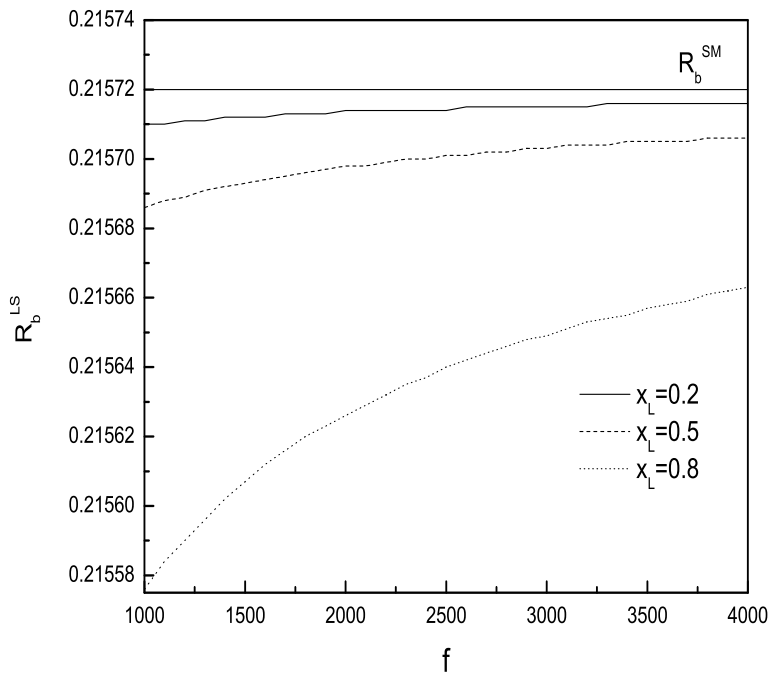,width=320pt,height=300pt} \vspace{-1cm}
\hspace{5mm} \caption{The branching ratio $R_{b}^{LS}$ as a
function of the scale parameter $f$ for three values of the mixing
parameter $x_{L}$.}
 \label{ee}
\end{center}
\end{figure}

Certainly, the neutral scalars $H^{0}$ and $\Phi^{0}$ can also
generate corrections to the $Z b\overline{b}$ couplings
$g_{L}^{b}$ and $g_{R}^{b}$ via the couplings $H^{0}b\overline{b}$
and $\Phi^{0}b\overline{b}$. However, compared with the charged
scalar contributions, the neutral scalar contributions are
suppressed at least by the factor $\frac{m_{b}^{2}}{m_{t}^{2}}$
and thus can be ignored. In order to get a positive definite mass
$M_{\Phi}$ of the triplet scalars, we should have that the value
of the ratio of the scalar triplet VEV $\nu'$ to the scalar
doublet VEV $\nu$ is smaller than $\frac{\nu}{4f}$. To simply our
calculation, we assume
$\frac{\nu'}{\nu}=\frac{1}{5}\frac{\nu}{f}$. In this case, the
triplet scalar mass $M_{\Phi}$ given in appendix A can be written
as: $M_{\Phi}=10m^{2}_{H}f^{2}/\nu^{2}$, which $m_{H}$ is the SM
Higg mass.

\begin{figure}[htb]
\vspace{-0.5cm}
\begin{center}
\epsfig{file=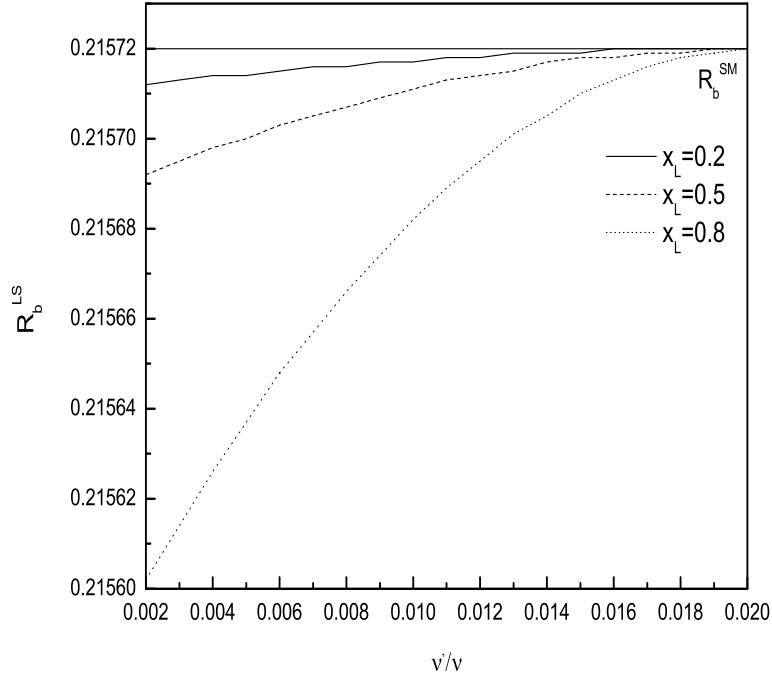,width=320pt,height=300pt} \vspace{-1cm}
\hspace{5mm} \caption{The branching ratio $R_{b}^{LS}$ as a
function of $\frac{\nu'}{\nu}$ for $f=3TeV$ and $x_{L}=0.2, 0.5$
and $0.8$.}
 \label{ee}
\end{center}
\end{figure}

In Fig.8 we plot the branching ratio $R_{b}^{LS}=R_{b}^{SM}+\delta
R_{b}^{LS} $ as a function of the scale parameter $f$ for
$x_{L}=0.2$(solid line), $0.5$(dashed line), and $0.8$(dotted
line), in which we have taken the SM Higgs mass $m_{H}=120GeV$. We
can see from Fig.8 that the charged scalars $\Phi^{\pm}$ generate
the negative corrections to the branching ratio $R_{b}$. The
negative corrections increase as the scale parameter $f$
decreasing and the mixing parameter $x_{L}$ increasing. For the
parameter $f\rightarrow \infty$, the corrections of the charged
scalars to $R_{b}$ go to zero. However, the varying value of
$R_{b}$ is very small and is smaller than that generated by the
new gauge bosons, the top quark $t$ and vector-like quark $T$ in
most of the parameter space.

To see the effects of varying the triplet scalar VEV $\nu'$ on the
branching ratio $R_{b}$, we take $f=3TeV$, which means
$\frac{\nu'}{\nu}<\frac{\nu}{4f}=0.0205$. The $R_{b}^{LS}$ is
plotted in Fig.9 as a function of $\frac{\nu'}{\nu}$ for $f=3TeV$
and three values of the mixing parameter $x_{L}$. From Fig.9 we
can see that the contributions of the charged scalars to $R_{b}$
decrease as the value of the ratio $\frac{\nu'}{\nu}$ increasing
for the fixed value of the parameter $f$ and
$\frac{\nu'}{\nu}<\frac{\nu}{4f}$. If we assume that the value of
the ratio $\frac{\nu'}{\nu}$ goes to $\frac{\nu}{4f}$, then the
correction value goes to zero.

\vspace{.5cm} \noindent{\bf VI. Discussions and conclusions}

The LH model predicts the existence of several scalars, new gauge
bosons, and vector-like quark $T$. These new particles can
generate corrections to the branching ratio $R_{b}$. Thus, the
predicted value of $R_{b}$ can be written as
$R_{b}^{LH}=R_{b}^{SM}+\delta R_{b}^{LG}+\delta R_{b}^{LT}+\delta
R_{b}^{LS}$ in the LH model. So, using the experimental value
$R_{b}^{exp}$, we might give the constraints on the free
parameters of the LH model.

\begin{figure}[htb]
\vspace{-0.5cm}
\begin{center}
\epsfig{file=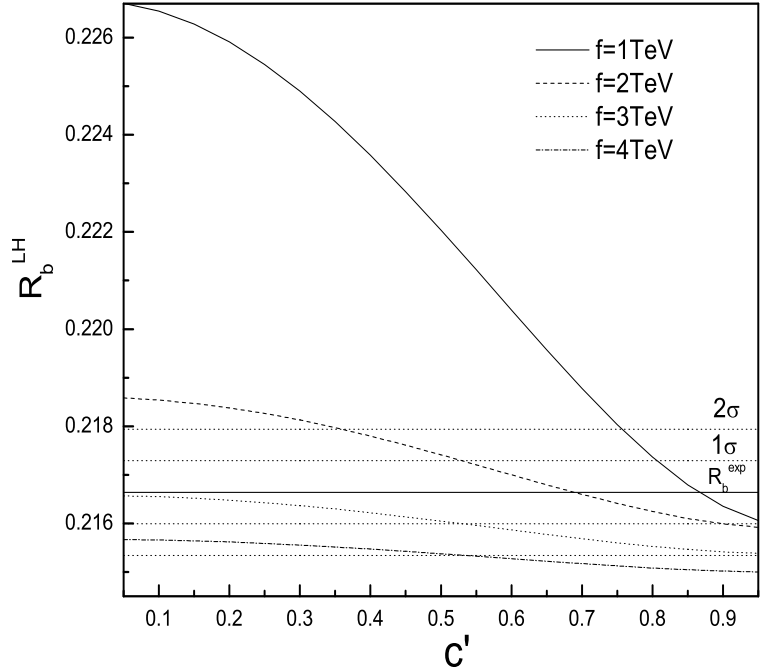,width=320pt,height=300pt} \vspace{-1cm}
\hspace{5mm} \caption{The predicted value of $R_{b}^{LH}$ in the
LH model as a function of the mixing parameter $c'$ for four
values of the scale parameter $f$.}
 \label{ee}
\end{center}
\end{figure}

From above discussions one can see that the correction effects of
the new particles predicted by the LH model to the branching ratio
$R_{b}$ decrease as the scale parameter $f$ increasing. The
charged scalars generate the negative correction to $R_{b}$ in all
of the parameter space. The correction value increases as the
mixing parameter $x_{L}$ increasing, which is very small. The
contributions of the top quark $t$ and the vector-like $T$ are
related to the parameters $c$ and $x_{L}$. However, they are
insensitive to the parameter $c$, while are strongly dependent on
the parameters $x_{L}$ and $f$. The new gauge bosons, such as $Z'$
and $B'$, can give corrections to $R_{b}$ at tree-level and
one-loop. The one-loop contributions are smaller than the
tree-level contributions at least by two orders of magnitude in
most of the parameter space. These contributions are sensitive to
the parameters $c'$ and $f$. Thus, the total correction of the LH
model to the branching ratio $R_{b}$ is mainly dependent on the
parameters $f$, $c'$ and $x_{L}$. Thus, we can take the parameters
$c$ and $\frac{\nu'}{\nu}$ as fixed value: $c=\frac{1}{\sqrt{2}}$
and $\frac{\nu'}{\nu}=\frac{\nu}{5f}$ for calculating the total
correction to $R_{b}$.

\begin{figure}[htb]
\vspace{-0.5cm}
\begin{center}
\epsfig{file=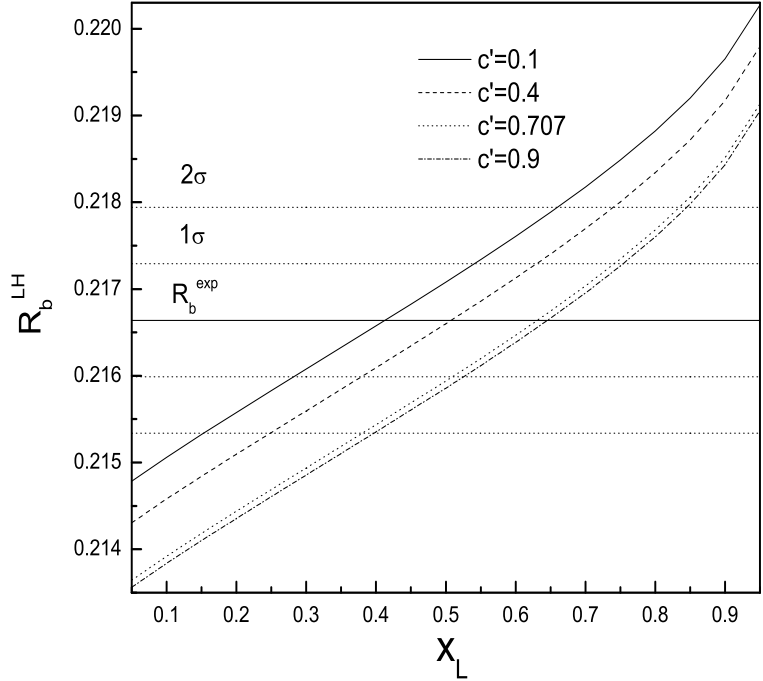,width=320pt,height=300pt} \vspace{-1cm}
\hspace{5mm} \caption{The predicted value of $R_{b}^{LH}$ in the
LH model as a function of the mixing parameter $x_{L}$ for four
values of the mixing parameter $c'$.}
 \label{ee}
\end{center}
\end{figure}

In Fig.10 we plot the branching ratio $R_{b}^{LH}$ as a function
of the mixing parameter $c'$ for $x_{L}=0.5$ and four values of
the scale parameter $f$. From Fig.10 we can see that the value of
$R_{b}^{LH}$ decreases as the parameter $c'$ increasing. For
$f=1TeV$, the value of $R_{b}$ is too large to consistent with the
precision experimental value $R_{b}^{exp}$ in most of the
parameter space. Furthermore, the large value of the scale
parameter $f$ is in favor of the general expectation based on
other phenomenological explorations. Thus, in Fig.11, we take
$f=3TeV$ and plot the $R_{b}^{LH}$ as a function of the mixing
parameter $x_{L}$ for  four values of the parameter $c'$. From
Fig.11 we can see that the value of $R^{LH}_{b}$ decreases as the
parameter $x_{L}$ increasing. If we demand that the predicted
value $R^{LH}_{b}$ consistent with the precision experimental
value $R^{exp}_{b}$ within $2\sigma$ bound for $f=3TeV$, there
must be:
\begin{eqnarray}
c'=0.1,\hspace{1cm}0.16\leq x_{L}\leq0.67;\hspace{1cm}
c'=0.4,\hspace{1cm}0.25 \leq x_{L}\leq 0.74;\nonumber\\
c'=\frac{1}{\sqrt{2}},\hspace{1cm}0.38\leq
x_{L}\leq0.84;\hspace{1cm} c'=0.9,\hspace{1cm} 0.39\leq
x_{L}\leq0.83.\nonumber
\end{eqnarray}
If we take the small value for the scale parameter f, these
constraints will became more strong. For example, for $f=2TeV$ and
$c'=c=\frac{1}{\sqrt{2}}$, we have $0.28\leq x_{L}\leq 0.65$ in
order to $R_{b}^{LH}$ consistent with $R_{b}^{exp}$ within
$2\sigma$ bound.

Little Higgs  models have generated much interest as possible
alternatives to weak scale supersymmetry. The LH model is a
minimal model of this type, which realizes the little Higgs idea.
In this paper, we study the corrections of the new particles
predicted by the LH model to the branching ratio $R_{b}$. We find
that the corrections of the neutral scalars to $R_{b}$ is very
small, which can be neglected. The charged scalars can generated
the negative corrections to $R_{b}$. The new gauge bosons and
fermions might generate the positive or negative corrections to
$R_{b}$, which dependent on the values of the mixing parameters
$c$, $c'$ and $x_{L}$. If we demand that the contributions of the
new gauge bosons and fermions cancel those generated by the
charged scalars and make the predicted value $R_{b}^{LH}$
consistent with the precision experimental value $R_{b}^{exp}$,
then the parameters $x_{L}$, $c'$ and $f$ must be severe
constrained.

\vspace{1.5cm} \noindent{\bf Acknowledgments}

This work was supported in part by the National Natural Science
Foundation of China (90203005).

\newpage
\vspace{.5cm} \noindent{\bf Appendix A: The masses of the gauge
bosons $Z'$ and $B'$, triplet scalar $\Phi$, and the vector-like
quark $T$.}

The masses of the gauge bosons $Z'$, $B'$ and $W'$ can be written
at the order of $\frac{\nu^{2}}{f^{2}}$:
\begin{eqnarray}
M_{Z'}^{2}&=&m_{Z}^{2}C_{W}^{2}[\frac{f^{2}}{s^{2}c^{2}\nu^{2}}-1-\frac{5S_{W}^{3}}{2C_{W}}\cdot
\frac{sc(c^{2}s'^{2}+s^{2}c'^{2})}{s'c'(5C_{W}^{2}s'^{2}c'^{2}-S_{W}^{2}s^{2}c^{2})}],\\
M_{B'}^{2}&=&m_{Z}^{2}S_{W}^{2}[\frac{f^{2}}{5s'^{2}c'^{2}\nu^{2}}-1+\frac{5C_{W}^{3}}{8S_{W}}\cdot
\frac{s'c'(c^{2}s'^{2}+s^{2}c'^{2})}{sc(5C_{W}^{2}s'^{2}c'^{2}-S_{W}^{2}s^{2}c^{2})}],\\
M_{W'}^{2}&=&m_{Z}^{2}c_{W}^{2}(\frac{f^{2}}{s^{2}c^{2}\nu^{2}}-1),
\end{eqnarray}
where $m_{Z}$ is the mass of the SM gauge boson $Z$, $\nu$ is the
electroweak scale.

For  the triplet scalar $\Phi$, we have
\begin{equation}
M_{\Phi}^{2}=2m_{H}^{2}\frac{f^{2}}{\nu^{2}}\frac{1}{1-(\frac{4f}{\nu}\frac{\nu'}{\nu})^{2}},\nonumber
\end{equation}
where $m_{H}$ is the SM Higgs mass and $\nu^{'}$ is the triplet
scalar vacuum expectation value(VEV).

The mass of the heavy vector-like quark $T$ can be written as:
\begin{equation}
M_{T}=\frac{m_{t}f}{\nu}\sqrt{\frac{1}{x_{L}(1-x_{L})}}[1-\frac{\nu^{2}}{2f^{2}}x_{L}(1+x_{L})],
\end{equation}
where $m_{t}$ is the SM top quark mass, $x_{L}$ is the mixing
parameter between the SM top quark and the heavy vector-like quark
$T$, which is defined as
$x_{L}=\frac{\lambda_{1}^{2}}{\lambda^{2}_{1}+\lambda^{2}_{2}}$.
$\lambda_{1}$ and $\lambda_{2}$ are the Yukawa coupling
parameters.

\vspace{.5cm}
\noindent{\bf Appendix B: The relevant coupling
constants of the gauge bosons to fermions.}
\begin{eqnarray}
W\overline{t}b:
g_{L}^{\overline{t}b}&=&\frac{ie}{\sqrt{2}S_{W}}[1-\frac{\nu^{2}}{2f^{2}}
(x_{L}^{2}+c^{2}(c^{2}-s^{2}))]V_{tb}^{SM},\\
g_{R}^{\overline{t}b}&=&0,
\end{eqnarray}
where $V_{tb}^{SM}$ is the SM CKM matrix element. In our
calculation, we will take $V_{tb}^{SM}=1$.

\begin{eqnarray}
&W\overline{T}b:
&g_{L}^{\overline{T}b}=\frac{e}{\sqrt{2}S_{W}}\frac{\nu}{f}x_{L}V_{tb}^{SM},
\hspace{1cm}g_{R}^{\overline{T}b}=0.\\
&W'\overline{t}b:
&g_{L}^{\overline{t}b}=-\frac{e}{\sqrt{2}S_{W}}\frac{c}{s}V_{tb}^{SM},
\hspace{1cm}g_{R}^{\overline{t}b}=0.\\
&W'\overline{T}b:
&g_{L}^{\overline{t}b}=-\frac{e}{\sqrt{2}S_{W}}\frac{\nu}{f}x_{L}\frac{c}{s}
V_{tb}^{SM},\hspace{1cm}g_{R}^{\overline{t}b}=0.\\
&Zb\overline{b}:
&g_{L}^{b}=\frac{e}{S_{W}C_{W}}\{-\frac{1}{2}+\frac{1}{3}S_{W}^{2}+\frac{\nu^{2}}
{f^{2}}[\frac{c^{2}(c^{2}-s^{2})}{4}\\
&&\hspace{1.5cm}+\frac{5}{6}(c'^{2}-s'^{2})(\frac{1}{5}-\frac{1}{2}c'^{2})]\},\nonumber\\
&&g_{R}^{b}=\frac{e}{S_{W}C_{W}}[\frac{1}{3}S_{W}^{2}+\frac{5}{3}\frac{\nu^{2}}
{f^{2}}(c'^{2}-s'^{2})(\frac{1}{5}-\frac{1}{2}c'^{2})].\nonumber\\
&Zt\overline{t}:
&g_{L}^{t}=\frac{e}{S_{W}C_{W}}\{1-\frac{2}{3}S_{W}^{2}-\frac{\nu^{2}}{f^{2}}
[x_{L}^{2}+\frac{c^{2}(c^{2}-s^{2})}{4}\\
&&\hspace{1.5cm}+\frac{5}{2}(c'^{2}-s'^{2})(\frac{4}{5}-c'^{2}+\frac{2}{3}s'^{2}x_{L})]\},\nonumber\\
&&g_{R}^{t}=\frac{e}{S_{W}C_{W}}\{-\frac{2}{3}S_{W}^{2}-\frac{\nu^{2}}
{f^{2}}5(c'^{2}-s'^{2})[\frac{3}{5}-c'^{2}(1-\frac{1}{3}x_{L}+\frac{2}{15}x_{L})]\}.\nonumber\\
&ZT\overline{T}: &g_{L}^{T}\approx
g_{R}^{T}=\frac{e}{S_{W}C_{W}}(-\frac{2}{3}S_{W}^{2}),\\
&Zt\overline{T}: &g_{L}^{tT}=-i\frac{e}{S_{W}C_{W}}\frac{x_{L}\nu}{4f},\hspace{1cm}g_{R}^{tT}=0.\\
&ZW^{+}W^{-}: &g^{ZWW}=g^{ZW'W'}=-\frac{eC_{W}}{S_{W}}\\
&Z'b\overline{b}: &g_{L}^{b}=-\frac{e}{2S_{W}}\cdot
\frac{c}{s},\hspace{1cm}g_{R}^{b}=0\\
&Z't\overline{t}: &g_{L}^{t}=\frac{e}{S_{W}}\cdot
\frac{c}{2s},\hspace{1cm} g_{R}^{t}=0. \\
&Z'T\overline{T}:
&g_{L}^{T}=g_{R}^{T}=-\frac{e}{S_{W}C_{W}}(\frac{2}{3}S_{W}^{2}).\\
&B'b\overline{b}:
&g_{L}^{b}=\frac{e}{3C_{W}s'c'}(\frac{1}{5}-\frac{1}{2}c'^{2}),\hspace{1cm}g_{R}^{b}
=\frac{2e}{3C_{W}s'c'}(-\frac{1}{5}+\frac{1}{2}c'^{2}).\\
&B'c\overline{c}:
&g_{L}^{c}=\frac{e}{3C_{W}s'c'}(\frac{1}{5}-\frac{1}{2}c'^{2}),\hspace{1cm}g_{R}^{c}
=\frac{4e}{3C_{W}s'c'}(\frac{1}{5}-\frac{1}{2}c'^{2}).
\end{eqnarray}

\vspace{.5cm}
\noindent{\bf Appendix C: The coupling constants of
the scalars to fermions.}

\begin{eqnarray}
H^{0}b\overline{b}:&&-i\frac{m_{b}}{\nu}(1-4\frac{\nu'^{2}}{\nu^{2}}+
2\frac{\nu'}{f}-\frac{2}{3}\frac{\nu^{2}}{f^{2}}).\\
\Phi^{0}b\overline{b}:&&-i\frac{m_{b}}{\sqrt{2}\nu}(\frac{\nu}{f}-4\frac{\nu'}{\nu}).\\
\Phi^{P}b\overline{b}:&&\frac{m_{b}}{\sqrt{2}\nu}(\frac{\nu}{f}-4\frac{\nu'}{\nu}).\\
\Phi^{+}\overline{t}b:&&-\frac{i}{\sqrt{2}\nu}[m_{t}P_{L}+m_{b}P_{R}](\frac{\nu}{f}-4\frac{\nu'}{\nu}),
\end{eqnarray}
with
$P_{L}=\frac{1-\gamma_{5}}{2}$,\hspace{1cm}$P_{R}=\frac{1+\gamma_{5}}{2}$.
\begin{equation}
\Phi^{+}\overline{T}b:\hspace{0.5cm}-i\frac{m_{t}}{\sqrt{2}\nu}(\frac{\nu}{f}
-4\frac{\nu'}{\nu})\sqrt{\frac{x_{L}}{1-x_{L}}}P_{L}.
\end{equation}

The coupling vertex of the SM gauge boson $Z$ to the charged
scalars $\Phi^{\pm}$ is
$$i\frac{e}{S_{W}C_{W}}S_{W}^{2}(P_{1}-P_{2})_{\mu}.$$


\newpage
\begin{thebibliography}{99}

\bibitem{1}
        D. Comelli and J. P. Silva, {\em Phys. Rev. D}{\bf
        54}(1996)1176; V. D. Barger, K. M. Cheung, P. Langaclar,
        {\em Phys. Lett. B}{\bf 381}(1996)226; P. Bamert, C. P.
        Burgess, J. M. Cline, D. London, and E. Nardi, {\em Phys. Rev. D}{\bf
        54}(1996)4275; D. Atwood, L. Reina, and A. Soni, {\em Phys. Rev. D}{\bf
        54}(1996)3296.
\bibitem{2}
        N. Arkani-Hamed, A. G. Cohen, E. Katz, A. E.
        Nelson, {\em hep-ph}/{\bf0206021}.
\bibitem{3}
         N. Arkani-Hamed, A. G. Cohen and H. Georgi,
         {\em  Phys. Lett. B}{\bf 513}(2001)232; N. Arkani-Hamed, A. G. Cohen,
         T. Gregoire and J. G. Wacker,
         {\em hep-ph}/{\bf0202089}; N. Arkani-Hamed, A. G. Cohen,
         E. Katz, A. E.  Nelson, T. Gregoire and J. G. Wacker,
         {\em hep-ph}/{\bf0206020}; I. Low, W. Skiba and
         D. Smith, {\em Phys. Rev. D}{\bf 66}(2002)072001; M.
         Schmaltz, {\em Nucl. Phys. Proc. Suppl.} {\bf
         117}(2003)40; D. E. Kaplan and M. Schmaltz,
         {\em hep-ph}/{\bf0302049}.
\bibitem{4}
         J. G. Wacker, {\em hep-ph}/{\bf0208235}; S. Chang and J. G. Wacker,
         {\em hep-ph}/{\bf0303001};
         W. Skiba and J. Terning, {\em Phys. Rev. D}{\bf 68}(2003)075001; S.
         Chang, {\em hep-ph}/{\bf0306034}.
\bibitem{5}
        T. Han, H. E. Logan, B. McElrath and L. T. Wang,
        {\em Phys. Rev. D}{\bf 67}(2003)095004.
\bibitem{6}
        M. Perelstein, M. E. Peskin and A. Pierce, {\em hep-ph} / {\bf0310039}.
\bibitem{7}
        G. Burdman. M. Perelstein and A. Pierce,
        {\em Phys. Rev. Lett.} {\bf 90}(2003)241802;
        C. Dib, R. Rosenfeld and A. Zerwekh, {\em hep-ph}/{\bf0302068};
        T. Han, H. E. Logan, B. McElrath ans L. T. Wang,
        {\em Phys. Lett. B}{\bf 563}(2003)191; Z. Sullivan,
        {\em hep-ph}/{\bf0306266}; S. C. Park and J. Song, {\em hep-ph}/{\bf0306112};
        Chongxing Yue, Shunzhi Wang, Dongqi Yu,
        {\em hep-ph}/{\bf0309113}.
\bibitem{8}
         C. Csaki, J. Hubisz, G. D. Kribs, P. Meade and J.
         Terning, {\em Phys. Rev. D}{\bf 67}(2003)115002;
         {\em D}{\bf68}(2003)035009; T. L. Hewett, F. J.
         Petrielo and J. G. Rizzo, {\em hep-ph}/{\bf 0211218}; T. Gregoire,
         D. R. Smith and J. G. Wacker, {\em hep-ph}/{\bf0305275};
         Wujun Huo and Shouhua Zhu, {\em Phys. Rev. D}{\bf 68}(2003)097301;;
         N. Mahajan, {\em hep-ph}/{\bf 0310098}.
\bibitem{9}
        Mu-Chun Chen and S. Dawson, {\em hep-ph}/{\bf 0311032}; R.
        Casalbuoni, A. Deandrea, M. Oertel, {\em
        hep-ph}/{\bf 0311038}; W. Kilian and J. Reuter, {\em hep-ph}/{\bf
        0311095}; S. Chang and Hong-Jian He, {\em
        hep-ph}/{\bf0311177}; C. Kilic and R. Mahbubani, {\em
        hep-ph}/{\bf0312053}.
\bibitem{10}
         S. R. Coleman and F. Weinberg, {\em Phys. Rev.D}{\bf 7}(1973)1888.
\bibitem{11}
         For review see V. A. Novikov, L. B. Okun, A. N. Rozanov, and M. I.
         Vysotsky, {\em Rept. Prog. Phys.} {\bf 62}(1999)1275.
\bibitem{12}
        J. Bernabeu, A. Pich, A. Santamaria, {\em Nucl. Phys. B}{\bf
        363}(1991)326; A. Denner, R. J. Guth, W. Hollik and J. H.
        Kuhn, {\em Z. Phys. C}{\bf 51}(1991)695; A. K. Grant, {\em Phys. Rev. D}{\bf
        51}(1995)207.
\bibitem{13}
        D. E. Groom, et al. [Particle Data Group], {\em Eur. Phys. J.
        C}{\bf 15}(2000)1; K. Hagiwora et al. [Particle Data
        Group], {\em Phys. Rev. D}{\bf 66}(2002)010001.
\bibitem{14}
        C. T. Hill and X. Zhang, {\em Phys. Rev. D}{\bf
        51}(1995)3563; C. X. Yue, Y. P. Kuang, X. L. Wang, and W.
        B. Li, {\em Phys. Rev. D}{\bf 62}(2000)055005.
\bibitem{15}
        C. X. Yue, Y. P. Kuang, G. R. Lu and L. D. Wan, {\em Phys. Rev. D}{\bf
        52}(1995)5314; C. X. Yue, Y. B. Dai and H. Li, {\em Mod. Phys. Lett. A}{\bf
        17}(2002)261.
\bibitem{16}
        P. Langacker, {\em hep-ph}/{\bf 0308145}; P. Gambino, {\em hep-ph}/{\bf
        0311257}.
\bibitem{17}
        J. Bernabeu, A. Pich and A. Santamaria, {\em Phys. Lett. B}{\bf
        200}(1988)569.
\bibitem{18}
        J. A. Aguilar-Saavedra, {\em Phys. Rev. D}{\bf 67}(2003)035003.
\bibitem{19}
        G. Passarino and M. Veltman, {\em Nucl. Phys. B}{\bf
        160}(1979)151; A. Axelrod, {\em Nucl. Phys. B}{\bf
        209}(1982)349; M. Clements etal., {\em Phys. Rev. D}{\bf 27}(1983)570.
\end{thebibliography}
\end{document}